\documentclass[usenatbib]{mn2e}
\usepackage{graphicx}
\begin{document}

\title[Fundamental parameters of RR Lyrae stars from multicolour
photometry]
{Fundamental parameters of RR Lyrae stars from multicolour photometry
and Kurucz atmospheric models. I. Theory and practical implementation}
\author[S. Barcza]
{S. Barcza$^1$\thanks{E-mail: barcza@konkoly.hu} \\
$^1$Konkoly Observatory, PO Box 67,
                   1525 Budapest, XII,
                   Hungary
}
\date{}

\maketitle

\begin{abstract} A photometric calibration of Kurucz static
  model atmospheres is used to obtain the following parameters of RR Lyrae 
  stars: variation of stellar angular radius $\vartheta$, effective
  temperature $T_{\rm e}$, gravity $g_{\rm e}$ as a function of
  phase, interstellar reddening $E(B-V)$ towards the star and
  atmospheric metallicity $M$. Photometric and hydrodynamic conditions
  are given to find the phases of pulsation when the quasi-static
  atmosphere approximation (QSAA) can be applied. The QSAA is
  generalized to a non-uniformly moving spherical atmosphere, and the
  distance $d$, mass ${\cal M}$ and atmospheric motion are
  derived from the laws of mass and momentum conservation. To
  demonstrate the efficiency of the method, the $UBV(RI)_C$
  photometry of SU Dra was used to derive the following
  parameters: $[M]=-1.60\pm .10$~dex, $E(B-V)=0.015\pm .010$,
  $d=663\pm 67$~pc, ${\cal M}=(0.68\pm .03){\cal M}_\odot$,
  equilibrium luminosity $L_{\rm eq}=45.9\pm 9.3L_\odot$ and 
  $T_{\rm eq}=6813\pm 20$~K.
\end{abstract}
\begin{keywords} hydrodynamics 
-- stars: atmospheres 
-- stars: fundamental parameters  
-- stars: individual: SU Dra
-- stars: oscillations 
-- stars: variables: RR Lyrae.   
\end{keywords}

\section{Introduction}

Although the determination of the fundamental parameters
of RR Lyrae (RR) stars is interesting in itself, it is also 
important from a practical point of view because
RR stars play a considerable role in establishing Galactic and
extragalactic distance scales. The Preston index and
spectroscopic observations are used to determine their atmospheric
metallicity $[M]$ and interstellar reddening $E(B-V)$. Since
confirmed RR-type components are not known in binary systems,
the mass determination is based on both stellar evolution and
pulsation theories. Due to the uncertainty of parallax data,
the Baade-Wesselink (BW) method is mostly used to infer
their distance $d$ \citep{smit1}. In the BW analysis, and to 
determine $[M]$ and $E(B-V)$, the quasi-static atmosphere
approximation (QSAA) is employed to interpret photometry and
spectroscopy.

The QSAA was introduced by \citet{ledo1}: `The simplest approach
  is to assume that at each phase, the atmosphere adjusts itself
  practically instantaneously to the radiative flux coming from the
  interior and to the effective gravity $g_{\rm e}$
\begin{equation}\label{1.102}
  g_{\rm e}=G{\cal M}R^{-2}+{\ddot R}
\end{equation}
where $R,\: \mbox{and}\: {\ddot R}$ are the instantaneous values of
the radius and acceleration, which is supposed uniform throughout the
atmosphere', $G$ is the Newtonian gravitational
constant, ${\cal M}$ is the stellar mass and dot is a
differentiation with respect to time $t$.  `One may then build a
  series of static model atmospheres,' and select one of them at
each phase by spectroscopic or photometric observations. Its flux,
colours, effective temperature $T_{\rm e}$ and surface gravity
$g_{\rm e}$ are accepted as the atmospheric parameters of that phase
providing a basis for the determination of other parameters such as angular
radius, mass, distance, etc.

The subject of the this paper is the QSAA and its 
generalization to a non-uniform atmosphere. We will investigate the
QSAA from the point of view of atmospheric emergent flux and
hydrodynamics. By comparing the observed colour indices with
those of static model atmospheres (\citealt{cast1}, \citealt{kuru1})
we select the phases when they coincide. Considering
hydrodynamics, we do not construct a consistent dynamic model of
an RR atmosphere. However, we find a better description of the
pulsating atmosphere if we characterize it by pressure and density
stratifications in addition to the two parameters $R$ and ${\ddot R}$.
We determine the fundamental parameters 
${\cal M}$ and $d$ 
from the hydrodynamic considerations without using the BW method
or theories of stellar evolution and pulsation at all. Our method 
uses photometry as observational input; spectroscopy and
radial velocity observations are not needed.

In Section 2, conditions of the QSAA are formulated for a
spherically pulsating compressible stellar atmosphere with a velocity
gradient. Practical methods are described to determine $[M], E(B-V)$,
$T_{\rm e}(\varphi)$ and $\log g_{\rm e}(\varphi)$ from $UBV(RI)_C$
colours of Kurucz atmospheric models, $\varphi$ being the
pulsation phase. The laws of mass and momentum conservation are used
to determine the mass and distance of the pulsating star.
Section 3 presents the results obtained from the $UBV(RI)_C$
photometry of SU Dra. The discussion and conclusions are given in Sections
4 and 5, respectively.

\section{The quasi-static atmosphere approximation}

First, we describe a photometric method to select the phases in
which QSAA can be regarded as a good approximation from a point
of view of fluxes. Next, we introduce the time-dependent
pressure and density stratifications of the photometrically selected
static models in the laws of mass and momentum conservation to find
the phases in which QSAA can be regarded as a more or less good
approximation from a hydrodynamic point of view as well. Using
the phases of valid QSAA from photometric and hydrodynamic points of
view, we determine mass, distance and infer the internal motions of
the atmosphere with respect to the stellar radius $R$. In 
this paper, we define $R$ as where the optical depth is 
zero, since the emergent fluxes of a theoretical model are given
for 
$\tau=0$ 
in any photometric band. Of course, the zero
boundary condition is a mathematical idealization \citep{ledo1}. 
In the following, we tacitly assume that the density is not
zero at the boundary, i.e.
$\varrho(R) > 0$ if $0 \le \tau \ll 1$.

\subsection{QSAA from the photometric point of view}
\label{2.1}

The functions 
$T_{\rm e}[{\rm CI}_1,{\rm CI}_2,[M],E(B-V)]$ and
$g_{\rm e}[{\rm CI}_1,{\rm CI}_2,[M],E(B-V)]$ 
of the Kurucz atmospheric models form a suitable grid for 
interpolation in the ranges 
$6000 < T_{\rm e} < 8000$ and $1.5 < \log g_{\rm e} < 4.5$ 
if the colour indices 
${\rm CI}_1,{\rm CI}_2$ 
are selected appropriately from $UBV(RI)_C$ photometry. For fixed
$[M]$ and $E(B-V)$
the intersection of the two functions
$\{T_{\rm e}^{(i)}(\log g_{\rm e},{\rm CI}_1,{\rm CI}_2,
[M],E(B-V))\}_{i=1,2}$ 
as a function of 
$\log g_{\rm e}$
gives a pair of 
$T_{\rm e}(\varphi),\log g_{\rm e}(\varphi)$ 
for a given pair of 
${\rm CI}_1(\varphi),{\rm CI}_2(\varphi)$. 
Although it is more practical to use 
$U-2B+V$ 
instead of 
$U-B$ 
\citep{barc3}, the other common colour
indices, e.g.  
$U-V$ 
etc., will also be kept.

Using 
$UBV(RI)_C$ 
photometry, we can construct four
independent colour indices 
${\rm CI}_{i=1,2,3,4}$, 
i.e.~${4\choose 2}=6$ 
combinations of colour indices resulting in six pairs of 
$T_{\rm e}(\varphi),\log g_{\rm e}(\varphi)$ 
for a given 
$\varphi$.  
Since the 
$U$ band covers the Balmer jump,
the indicator of 
$g_{\rm e}(\varphi)$, 
the combinations must contain at least one colour index 
containing 
$U$ 
photometry. 
${10\choose 2}-{6\choose 2}=30$ 
such combinations can be constructed. The average of the 30 pairs of 
$T_{\rm e}(\varphi),\log g_{\rm e}(\varphi)$ 
will be accepted for a phase
$\varphi$.

{\it Condition I.} If the scatters 
$\Delta T_{\rm e}(\varphi),\Delta\log g_{\rm e}(\varphi)$ 
of 
$T_{\rm e}(\varphi),\log g_{\rm e}(\varphi)$ 
are compatible with the expected scatter from the error of 
the colour indices, QSAA provides a good approximation 
in this phase from the photometric point of view.

The bolometric corrections and 
$UBV(RI)_C$ 
physical fluxes of a model
with given 
$T_{\rm e},\log g_{\rm e},[M],E(B-V)$ 
are available in
tabular form \citep{kuru1} and they allow us to determine the angular
radius of the star defined by
\begin{equation}\label{2.202}
\vartheta(\varphi)=R(\varphi)/d.
\end{equation}
Technical details of finding the appropriate model and determining
$\vartheta$ 
at a phase 
$\varphi$ 
are given in \citet{barc1} and \citet{barc3}.\footnote{A 
program package to perform the conversion of colour indices and 
magnitudes to 
$T_{\rm e}$, $\log g$, $\vartheta$ 
is available upon request.}

On the scale provided by the Kurucz atmospheric models, the
above procedure offers a possibility to determine reddening and
atmospheric metallicity from photometry without using spectroscopy. 
In the shock free phases, the QSAA is expected to reproduce the
colours well, therefore, 
$\Delta T_{\rm e},\Delta\log g_{\rm e}$ 
must be minimal as a function of the 
$\varphi$-independent 
model parameters
$E(B-V),[M]$ \citep{barc3}. Of course the method can also be
applied for non-variable stars to determine 
$E(B-V)$, $[M]$, $T_{\rm e}$, $\log g$, $\vartheta$.

\subsection{QSAA from the hydrodynamic point of view}

In the original form of the QSAA \citep{ledo1}, the atmospheric
motion is described by the parameters 
$R(\varphi),{\dot R}(\varphi),{\ddot R}(\varphi)$ 
and the outward or inward
accelerations are driven by the variable 
$g_{\rm e}(\varphi)-g_{\rm s}[R(\varphi)]$ 
in (\ref{1.102}) where 
$g_{\rm s}=G{\cal{M}}/r^2$. 
This is essentially a uniform atmosphere
approximation (UAA). However, the atmosphere of an RR star is a
dilute, compressible gaseous system with variable temperatures. 
Therefore, the introduction of additional parameters containing 
$T_{\rm e}(\varphi)$ 
promises a better description in the frame of the QSAA.

Considering the geometry of radial pulsation modes \citep{smit1}
and neglecting rotation will result in all components of the
vector field velocity $v(r,t)$ being zero except for
the radial component $v(r,t)$. The viscosity is negligible at the
velocities occurring in an RR atmosphere. Therefore, the
momentum conservation is expressed by the Euler equation of
hydrodynamics:
\begin{equation}\label{1.100}
{{\partial v}\over{\partial t}}+v{{\partial v}\over{\partial r}}
+g_{\rm s}(r)
+{1\over{\hat\varrho}}{{\partial {\hat p}}\over{\partial r}}
+a^{\rm (tang)}(r,t)=0 
\end{equation}
where $t=P\varphi$, $P$ is the pulsation period.

The contribution of the neglected tangential motions is the average
$a^{\rm (tang)}=(4\pi)^{-1} \int_0^\pi {\rm d}\theta\sin\theta
\int_0^{2\pi}{\rm d}\phi [v_\theta(\partial v_r/\partial\theta)
+v_{\phi}\sin^{-1}\theta(\partial v_r/\partial\phi)
-v_{\theta}^2-v_{\phi}^2]/r$ \citep{land1}. In perfect spherical
symmetry $a^{\rm (tang)}=0$. The actual pressure and density
stratifications were devided into two parts 
${\hat p}(r,t)=p(r,t)+p^{\rm (dyn)}(r,t)$ and 
${\hat \varrho}(r,t)=\varrho(r,t)+\varrho^{\rm (dyn)}(r,t)$, 
where 
$p(r,t)$ and $\varrho(r,t)$ 
denote the values of a static model atmosphere, while 
$p^{\rm (dyn)}(r,t)$, $\varrho^{\rm (dyn)}(r,t)$ 
are the dynamical corrections due to pulsation.

From the hydrodynamic point of view, the essence of the QSAA is
that at a given $t$
\begin{equation}\label{2.100}
  -{1\over\varrho}{{\partial p}\over{\partial r}}=g_{\rm e}(t) > 0,
\end{equation}
i.e. the $r$-independent effective gravity of a static model
atmosphere is introduced in (\ref{1.100}) and 
$p^{\rm (dyn)}(r,t)=0$,
$\varrho^{\rm (dyn)}(r,t)=0$.  
In what follows, we drop this latter
restriction and define a dynamical correction term of acceleration
\begin{equation}\label{2.120}
  a^{\rm (dyn)}={1\over{{\hat\varrho}}}
  {{\partial {\hat p}}\over{\partial r}}
  -{1\over\varrho}{{\partial p}\over{\partial r}}
\end{equation}
to account for the difference of a static and dynamical atmosphere.
Because of the lack of dynamical model atmospheres, 
$a^{\rm (dyn)}$ 
must be determined empirically from the quantities derived from photometry.

Equation (\ref{1.100}) is transformed to definition  
(\ref{1.102}) of QSAA if 
$\partial v/\partial r = 0$ 
and 
$a^{\rm (tang)}=0$, 
$a^{\rm (dyn)}=0$, 
because these simplifications imply 
$\partial v/\partial t={\ddot R}$ 
and 
$g_{\rm s}(R)=G{\cal M}R^{-2}$, 
i.e. the UAA is valid. As a dynamical 
equation, this simplified equation of motion is the basis of 
mass determination in BW analyses (e.g. \citealt{cacc1}, \citealt{liuj1},etc).

If the sound velocity is assumed as an upper limit for 
$v_{\theta}$
and 
$v_{\phi}$, $a^{\rm (tang)}$ is some $0.1\mbox{ms}^{-2}$. 
It is negligible in comparison with the other components of the acceleration
at any 
$r \la R$.  
Typical values are 
$g_{\rm s} \la 10{\rm ms}^{-2}$
during the pulsation cycle of an RR star while 
$g_{\rm e}$ 
can exceed
$100{\rm ms}^{-2}$ 
when the atmosphere is in the state of
maximal compression. Thus, 
$a^{\rm (tang)}$ 
will be neglected and
\begin{equation}\label{2.101}
  g(r,t)=g_{\rm e}(t)-g_{\rm s}(r)
\end{equation}
is a periodic function of $t$ with zero-points.

If the $r$-dependence of the temperature $T$ is neglected and a
constant density upper boundary condition is assumed, an
integration of (\ref{2.100}) over the interval $[r,R]$ with the
equation of state of a perfect gas gives the approximate density
stratification of the model atmosphere:
\begin{equation}
\label{2.110}
\varrho(r,t)=\varrho(R)\exp\{-h_0(R,t)[r-R(t)]\}
\end{equation}
where $h_0(R,t)=\mu g_e(t)/{\cal R}T(R),{\cal R},\mu$ are the
reciprocal barometric scaleheight at 
$R$, 
the universal gas constant
and average molecular mass, respectively. Essentially, we use the
static model atmospheres to measure 
$-\varrho^{-1}(\partial p/\partial r)$. 
The atmospheric pulsation is driven by 
$g(r,t)-a^{\rm (dyn)}$ 
and the thermal
processes are represented by the variable 
$h_0$.

{\it Condition II.} If 
$\partial v/\partial t+v\partial v/\partial r\approx g$, 
i.e.~$\vert a^{\rm (dyn)}(r,t)\vert \la 2\Delta g$, 
QSAA provides a good approximation in this phase from the 
hydrodynamic point of view. 

This condition formulates the fact 
that the dynamical excess of acceleration in the upper
photosphere is smaller than the $\Delta g$ error of $g$ and the
same error is assumed for $\partial v/\partial t+v\partial v/\partial
r$.  The combination of (\ref{2.120}) and (\ref{1.100}) allows us to
estimate the constant and ${\rm O}(d)$ terms of $a^{\rm (dyn)}$.
However, the satisfaction of Condition II can be checked afterwards
when the terms with other powers of $d$ were determined. The quotient
$q(r,t)=a^{\rm (dyn)}/g$ characterizes the degree of excellence of
the QSAA. The QSAA is exact from the hydrodynamic point of view if 
$a^{\rm (dyn)}=0$.

The perfect spherical symmetry means that in a Euler picture
\citep{prin1} we have to use (\ref{1.100})-(\ref{2.110}) with 
$a^{\rm (tang)}(r,t)=0$ 
and the mass conservation law
\begin{equation}\label{1.101}
{{\partial\varrho}\over{\partial t}}
+v{{\partial\varrho}\over{\partial r}}
+\varrho{{\partial v}\over{\partial r}}
+{2\over r}v\varrho=0 
\end{equation}
must be taken into account.  With assumption (\ref{2.110}), the
analytic solution to (\ref{1.101}) is
\begin{equation}\label{2.302}
  v(r,t)=-a_1r+a_0+a_{-1}r^{-1}+a_{-2}r^{-2}
\end{equation}
where $a_1=Ch_0^{-1}(\partial h_0/\partial t)$, $a_0={\dot
  R}+Ca_1(R-3h_0^{-1})$, $a_{-1}=2Ca_0h_0^{-1}$,
$a_{-2}=Ca_{-1}h_0^{-1}$, $C=1$.  $C$ 
was introduced to incorporate UAA by taking 
$a^{\rm (dyn)}(R,t)=0$ and $C=0$. 
The main term in $v$ is ${\dot R}$ and the convergence of (\ref{2.302}) is
excellent because $rh_0\gg 1$.

After differentiations of $v$, $h_0$ and $\vartheta$, taking $r=R$,
the following form is convenient for a numerical solution:
\begin{equation}\label{107a}
{\cal M}(d,t_1)-{\cal M}(d,t_2)=0
\end{equation}
where 
${\cal M}(d,t)=[g_{\rm e}(t) -(\partial v/\partial t)
-v(\partial v/\partial r) -a^{\rm (dyn)}(R,t)]R^2/ G$.  It can be
solved if there exist two or more $t$ intervals satisfying Condition I
and 
$a^{\rm (dyn)}(R,t)$ 
is identical for them. The assumed
differences 
$a^{\rm (dyn)}(R,t_1)-a^{\rm (dyn)}(R,t_2)\approx 0$ 
and the satisfaction of Condition II must be verified afterwards. 
Equation (\ref{107a}) must be solved for different phase pairs 
$(\varphi_1,\varphi_2)$ 
and the values 
$d$ 
and 
${\cal M}$ must be omitted from the
final averaging if 
$a^{\rm (dyn)}(R,\varphi_1)\not\approx a^{\rm (dyn)}(R,\varphi_2)$.

To account for the difference of the local and effective
temperatures, the boundary temperature 
$T(R)=2^{-0.25}T_{\rm e}$
of a gray atmospheric model at $\tau=0$ \citep{miha1} will be used in
$h_0(R,t)$.  
Hydrogen and helium are mainly neutral at the boundary
temperature of RR stars. Consequently, assuming a typical
chemical composition means that $\mu=1.3$ is the appropriate choice in
$h_0(R,t)$.

\begin{figure}
  \includegraphics[width=84mm]{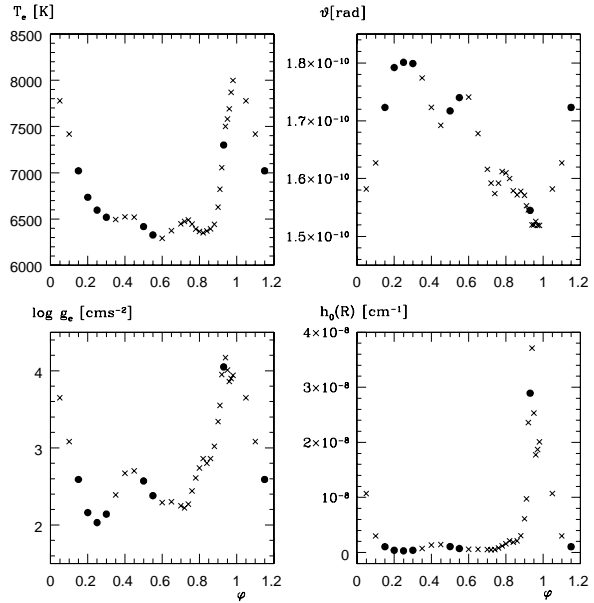}
\vspace{-0.5cm}
\caption{$T_{\rm e}$, $\vartheta$, $\log g_{\rm e}$, $h_0(R)$ of SU
  Dra as a function of phase. Filled circles indicate the phases when
  Condition I is satisfied.}
\label{fig1}
\end{figure}

\begin{figure}
  \includegraphics[width=84mm]{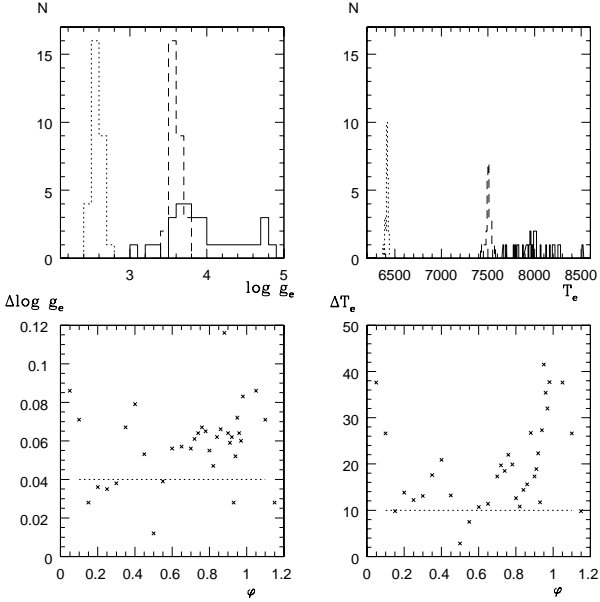}
\vspace{-0.5cm}
\caption{Upper panels: histogram of 30 $\log g_{\rm e}$, 
  $T_{\rm e}$ 
  values in bins of size $0.1,10\mbox{K}$, respectively. Dotted:
  SU Dra at $\varphi=0.5$, solid line: SU Dra at $\varphi=0.98$,
  dashed: BD +67\degr 708. Lower panels: standard errors 
  $\Delta\log g_{\rm e}$ and $\Delta T_{\rm e}$ 
  from 30 possible combinations of 
  ${\rm CI}_i$ 
  as a function of phase. At phases below the dotted
  lines the scatter can be attributed to observational errors of the
  colour indices.}
\label{fig2}
\end{figure}

\section{Results for SU Dra}

\citet{barc0} derived good quality light curves of the RRab star 
SU Dra by homogenizing 45 years of $UBV(RI)_C$ observations (see 
his table 7). The method described in the previous section
will now be applied. Instead of smoothing the light curves, 
binned data will be used to keep the results as close as possible to
observed quantities. The 30 colour index pairs containing at
least one $U$ were obtained from 
${\rm CI}_i\vert_{i=1}^{i=10}=
\{U-2B+V,U-V,U-R_C,U-I_C,B-V,B-R_C,B-I_C,V-R_C,V-I_C,R_C-I_C\}$.
Taking 
$[M]=-1.60$, $E(B-V)=0.015$ 
\citep{liuj1} the basic quantities
for (\ref{107a}) are summarized in Table~\ref{tab1} for different 
phases. Fig. 1 is a plot of 
$T_{\rm e}(\varphi)$,
$\vartheta(\varphi)$, $\log g_{\rm e}(\varphi)$, $h_0(R,\varphi)$ for
one whole pulsation.

$\Delta T_{\rm e}(\varphi)$ and $\Delta\log g_{\rm e}(\varphi)$ 
are plotted in the lower panels of Fig.~\ref{fig2}. Assuming
random errors of $\pm 0.02$ for the colour indices will 
result in 
$\Delta T_{\rm e}=\pm 10\mbox{K}$ 
and 
$\Delta\log g_{\rm e}=\pm 0.04$. 
These values are indicated by the dotted
horizontal lines. Condition I is satisfied in the phase points lying
below or close to the dotted lines. At phases lying above the
dotted lines, the monochromatic flux of static models and SU Dra
differs significantly on a level which has noticeable effect on the
broad band colours $UBV(RI)_C$. The assumed $[M]=-1.6$, $E(B-V)=0.015$
were verified by a variation procedure \citep{barc3}. In the shock
free phases $\varphi=0.15,0.5,0.55$, minimization of $\Delta
T_{\rm e}(\varphi),\Delta\log g_{\rm e}(\varphi)$ resulted in
$[M]=-1.60\pm 0.10$, $E(B-V)=0.015\pm 0.01$.

To demonstrate the difference between good and poor QSAA, the
histograms of the 
30 $\log g_{\rm e},T_{\rm e}$ 
values are plotted in the upper panels of Fig.~\ref{fig2} for 
$\varphi=0.5,0.98$ 
of SU Dra and BD +67 708. They show normal distributions with small
scatter for the non-variable BD +67\degr 708 and SU Dra at $\varphi=0.5$.
At 
$\varphi=0.98$, 
the distribution is almost uniform. At the next
phase point $\varphi=1$, merely 14 intersections of 
$\{T_{\rm e}^{(i)}(\log g_{\rm e},{\rm CI}_1,{\rm CI}_2,[M],E(B-V))\}_
{i=1,2}$, 
i.e.~14 pairs of 
$\log g_{\rm e},T_{\rm e}$ 
were found instead of 30 pairs. Thus, at 
$\varphi \approx 1$ 
the observed colours differ significantly from those of any static model
of \citet{kuru1}. However, it is interesting to note that the small
errors 
$\Delta\log g_{\rm e}=0.03$, $\Delta T_{\rm e}=12\mbox{K}$
indicate a phase island at 
$\varphi=0.93$ 
with observed and static model colours in agreement for some 
$0.01P\approx 10$~minutes. This
happens to be the phase of the hump on the light curve when the
inward and outward motions encounter and produce a shock
\citep{smit1}.

In the interval $0.92 < \varphi < 1.05$ the atmosphere is in a
state of maximal compression by the shock coming from the
sub-photospheric layers and $R$ is nearly minimal. This is the
rising branch and the start of the descending branch in the light 
curve. Therefore, the values of 
$\log g_{\rm e},T_{\rm e},\vartheta$ 
obtained from QSSA, if they can be found at all, must be considered 
as a first approximation only. This is reflected in large 
$\Delta \log g_{\rm e}$, $\Delta T_{\rm e}$ 
except for 
$\varphi \approx 0.93$.

\begin{table}
  \caption{The basic quantities for (\ref{107a}) at various phases. 
    The units are 
    ${\rm cms}^{-2}{\rm ,K,rad\times 10^{10},cm^{-1}\times 10^{10}}$.}
\label{tab1}
\begin{tabular}{lllllll}
\hline
$\varphi$ & $\log g_{\rm e}$ & $T_{\rm e}$ & $\vartheta$ 
                                             & $h_0(R)$ & \\
\hline
$0.1 $ & $3.08\pm .07$ & $7418\pm 27$ & $1.627\pm .004$ & $30.0	$&-,- \\
$0.15$ & $2.59\pm .03$ & $7021\pm 10$ & $1.723\pm .003$	& $10.4	$&I,II \\
$0.2 $ & $2.16\pm .04$ & $6735\pm 14$ & $1.792\pm .004$	& $3.99	$&I,- \\
$0.25$ & $2.03\pm .04$ & $6596\pm 12$ & $1.801\pm .004$	& $3.00	$&I,II \\
$0.3 $ & $2.14\pm .04$ & $6519\pm 13$ & $1.799\pm .004$	& $3.96	$&I,II \\
$0.35$ & $2.39\pm .07$ & $6495\pm 18$ & $1.774\pm .004$	& $7.10	$&-,II \\
$0.4 $ & $2.67\pm .08$ & $6524\pm 21$ & $1.723\pm .004$	& $13.4	$&-,- \\
$0.45$ & $2.70\pm .05$ & $6519\pm 13$ & $1.692\pm .004$	& $14.3	$&-,- \\
$0.5 $ & $2.57\pm .01$ & $6418\pm  3$ & $1.717\pm .003$	& $10.7	$&I,- \\
$0.55$ & $2.38\pm .04$ & $6328\pm  8$ & $1.740\pm .003$	& $7.11	$&I,II \\
$0.6 $ & $2.29\pm .06$ & $6292\pm 11$ & $1.741\pm .004$	& $5.83	$&-,- \\
$0.65$ & $2.30\pm .06$ & $6374\pm 11$ & $1.678\pm .004$	& $5.77	$&-,- \\
$0.7 $ & $2.25\pm .06$ & $6448\pm 17$ & $1.616\pm .006$	& $5.18	$&-,- \\
$0.92$ & $3.95\pm .06$ & $7055\pm 22$ & $1.546\pm .006$	& $235	$&-,- \\
$0.93$ & $4.05\pm .03$ & $7300\pm 12$ & $1.545\pm .003$	& $289	$&I,- \\
$0.94$ & $4.17\pm .05$ & $7502\pm 27$ & $1.520\pm .006$ & $371	$&-,II \\
$0.95$ & $4.01\pm .07$ & $7582\pm 42$ & $1.520\pm .008$ & $164	$&-,- \\
\hline
\end{tabular}

\smallskip
{\it Note.}
Satisfaction of Conditions I and/or II is indicated in the last column
by I and/or II, respectively.
\end{table}

\begin{figure}
  \includegraphics[width=84mm]{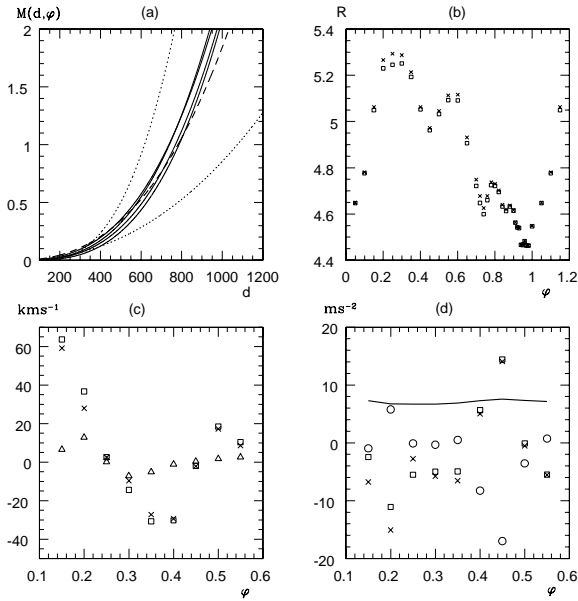}
\vspace{-0.5cm}
\caption{Panel (a): The functions ${\cal M}(d,\varphi)$ in units
  of solar mass and pc, respectively.  Dotted, from left to right:
  $\varphi=0.2,0.5$, dashed: $\varphi=0.25$, lines: from left to right
  $\varphi=0.55,0.35,0.3,0.15$. Panel (b): crosses:
  $R(\varphi)$, squares: $R(\varphi)-1/h_0(R)$, in solar units.
  Panel (c): crosses: ${\dot R(\varphi)}$, squares:
  $v(\varphi)$, triangles: $h_0^{-1}(\partial v/\partial
  r)\vert_{r=R}$. Panel (d): crosses: ${\ddot R(\varphi)}$,
  squares: $\partial v/\partial t+v\partial v/\partial r$, 
  circles: $a^{\rm (dyn)}(R,\varphi)$, thick line: $g_{\rm s}(R)$.  In
  panels (b)-(d) the angular variables were converted to absolute
  radius etc. by $d=663$~pc and ${\cal M}=0.68{\cal M}_\odot$.  }
\label{fig3}
\end{figure}
 
To obtain
$v(r,t)$, 
etc., for (\ref{107a}), 
$\vartheta(t)$ and $h_0(R,t)$
were differentiated by midpoint formulae. Fig.~\ref{fig3}(a) is a plot
of the functions 
${\cal M}(d,t)$ 
for the phases 
$t=\varphi P$,
$\varphi=0.15\mbox{-}0.35,0.5,0.55$ 
with 
$a^{\rm (dyn)}(R,t)=0$. 
The average and standard error of 
${\cal M},d$ 
are given in Table~\ref{tab2} from the pairs 
$(\varphi_1=0.25$ and $\varphi_2=0.15,0.3,0.35,0.55)$, 
$(\varphi_1=0.35$ and $\varphi_2=0.55)$ 
in (\ref{107a}) as our best values denoted by
$[\ast]$.  At $\varphi=0.35$ Condition I is moderately violated, but
Condition II is satisfied and $a^{\rm (dyn)}(R,t)/g_{\rm
  s}(R,t)\approx -0.07$; therefore, this phase was included to obtain
$d,{\cal M}$ of $[\ast]$. Condition I is satisfied at
$\varphi=0.2,0.5$; however, these phases had to be excluded from the
mass and distance determination because of the large $a^{\rm
  (dyn)}(R,\varphi=0.2,0.5)=5.8,-3.3\:\mbox{ms}^{-2}$ and
$q=-0.52,38$, respectively.

Using our best solution for ${\cal M}$ and $d$, the radius
variation, velocities and the components of acceleration were
computed in physical units and are plotted in
Fig.~\ref{fig3}(b)-(d). Velocities and accelerations are 
plotted only for the phases of more or less good QSAA
($\varphi=0.15\mbox{-}0.3,0.5,0.55$), 
including the slightly shocked
phases $\varphi=0.35\mbox{-}0.45$.

At the phases $\varphi=0.15,0.25,0.3,0.35,0.55$ 
$a^{\rm (dyn)}(R,\varphi)=-79,8,-16,65,93\:\mbox{cms}^{-2}$ 
and Condition I is satisfied, 
$\vert a^{\rm (dyn)}(R,t)/g_{\rm s}(R,t)\vert < 0.13$,
i.e.~$a^{\rm (dyn)}(R,\varphi)$ 
is small in comparison with the other acceleration terms in (\ref{1.100}).  
$\vert q\vert\approx 0.1$ 
is expected from 
$\Delta \log g_{\rm e}=0.04$, $q(R,\varphi)=0.39,$
$0.02,$ $0.06,$ $-0.10,$ $-0.13$ 
were found. Thus, Condition II is indeed satisfied. The outlier value of
$0.39$ 
is produced by cancellation of 
$\partial v/\partial t\approx -30.5\:\mbox{ms}^{-2}$
and 
$v\partial v/\partial r\approx +28.2\:\mbox{ms}^{-2}$, 
i.e.~the free fall of the upper atmosphere is for a short time almost uniform
with constant velocity and vanishing acceleration
$-g(R,\varphi)\approx -2.3\:\mbox{ms}^{-2}$. Therefore, the density
stratification differs from (\ref{2.110}) and this is manifested in
the large $q$. However, this phase could also be used to 
get the best
$d$ 
and 
${\cal M}$ 
in Table \ref{tab2} because 
$\vert a^{\rm (dyn)} \vert < 1\:\mbox{ms}^{-2}$.

\begin{table}
  \caption{Distance and mass of SU Dra from (\ref{107a}).}
\label{tab2}
\begin{tabular}{llll}
  \hline
  $d[{\rm pc}]$ & ${\cal M}[{\cal M}_\odot]$ & Used phases & Remark\\
  \hline
  \bf $663\pm 67$ & $0.68\pm .03$ & $0.15,0.25\mbox{-}0.35,0.55$ & 
  $C=1,\:\:[\ast]$       \\
  $658\pm 104$ & $0.65\pm .03$ & $0.15,0.25,0.30,0.55$ & $C=1$        \\
  $618\pm 200$ & $0.54\pm .10$ & $0.20,0.30,0.50,0.55$ & $C=0$, UAA   \\
  \hline
\end{tabular}

\smallskip
{\it Note.}
$[\ast]$
indicates the values which we accept as the best ones.
\end{table}

In the interval 
$0.92 \leq \varphi \leq 0.95$, 
the acceleration is dominated by 
$g_{\rm e}\approx 89,112,148,102\:\mbox{ms}^{-2} \gg g_{\rm s}$ 
and the atmosphere is practically at a standstill. The values
$q=1.06,-1.68,-0.12,1.44$ 
and 
$a^{\rm (dyn)}(R,t)/g_{\rm s}(R,t)\approx 5,28,-2,6$ 
indicate 
$a^{\rm (dyn)}\approx 41,254,-19,55\:\mbox{ms}^{-2}$ 
and 
$\vert a^{\rm (dyn)} \vert \gg g_{\rm s}$. 
If 
$g_{\rm s}$ 
is neglected by setting 
${\cal M}(d,t)=0$
and the data of Table~\ref{tab1} are used with 
$a^{\rm (dyn)}=0$, 
the upper limit 
$d < 642\mbox{pc}$ 
is obtained from
$\varphi=0.94$. 
If this estimation is improved by 
$a^{\rm (dyn)}(R,\varphi=0.94)=-19\mbox{ms}^{-2}$ 
obtained for the distance
and mass $[\ast]$, 
the upper limit shifts to $d < 724\mbox{pc}$.
From the functions 
${\cal M}(724\:\mbox{pc},\varphi=0.15,0.25\mbox{-}0.3,0.55)$ 
the upper mass limit is 
${\cal M} < 0.84\pm .03\:{\cal M}_\odot$. 
These upper limits
are in accordance with the values $[\ast]$. However, they 
should be treated as indicative only, because they are sensitive to
the actual value of 
${\dot\vartheta},{\ddot\vartheta},\partial h_0/\partial t,
\partial^2 h_0/\partial t^2$. 
These values could be determined
by numerical differentiations from $\vartheta,h_0$ in 
$0.91 \leq \varphi \leq 0.96$ 
when the QSAA is not optimal to obtain all of them. Hence, the phase island 
$\varphi=0.93\mbox{-}0.94$ 
could not be
included in the mass and distance determination because a very
transient good QSAA is embedded in a strongly shocked interval and the
mass ${\cal M}$ appearing in $g_{\rm s}$ plays a secondary role in the
dynamics.

For the sake of completeness, it must be mentioned that the pair 
$(\varphi_1=0.15$ and $\varphi_2=0.3)$ 
in (\ref{107a}) gives 
$d=1485$~pc.
Adding this value to those involved in $[\ast]$, 
$d=(800\pm 148)$~pc 
and 
${\cal M}=(1.15\pm .05){\cal M}_\odot$ 
is obtained. This distance is closer to 
$d=900$~pc 
suggested by the
{\it Hipparcos} parallax $\pi=1.11\pm 1.15$~mas \citep{hipp1}.
However, this $1485$~pc was omitted because it is an outlier point
above the 
$3\sigma$ 
level, it exceeds the upper limit $724$~pc and the
inaccurate knowledge of 
$a^{\rm (dyn)}(R,0.15)-a^{\rm (dyn)}(R,0.3)$
makes it uncertain.

Close to the shocked phases (i.e. for $\varphi=0.1$, $0.4$, $0.45$,
  $0.7\mbox{-}0.76$, $0.84$, $0.86$, $0.91$, $0.94$, $0.97$), 
  negative stellar masses were obtained if 
$a^{\rm (dyn)}(R,\varphi)=0$. 
The condition 
${\cal M}(d,\varphi) > 0$ 
gives an upper limit for 
$a^{\rm (dyn)}(R,\varphi) < 0$ 
at these phases and shows clearly the
limitations of treating an RR atmosphere in the QSAA. In these phases the
characteristic value $\vert q\vert\approx 1$. The 
sharp peaks 
$q(R,\varphi=0.5)=38.8$ and $q(R,\varphi=0.65)=-13.8$ 
are produced by 
$\partial v/\partial t\approx -v\partial v/\partial r$,
i.e.~the atmosphere is almost completely free of acceleration.
$q(R,\varphi=0.05)=11.4$ 
is a result from the poor representation of
the density stratification by (\ref{2.110}), i.e.~the dynamical term
is large in 
$g_{\rm e}-g_{\rm s}-a^{\rm
  (dyn)}=(44.7-8.7-33.1)\:\mbox{ms}^{-2}$ 
producing 
$a^{\rm (dyn)}(R,\varphi) \approx 11(\partial v/\partial t+v\partial
v/\partial r)$. 
This is the start of the descending branch in
the light curve, when the atmosphere starts expanding rapidly.

The averaged surface gravity for the whole pulsation cycle is
$\langle\log g_{\rm e}\rangle=2.73$, 
which is in excellent agreement with 
$\langle\log g_{\rm e}\rangle=2.69,2.72$ 
from $uvby$ and $UBV(RI)_C$ photometry (\citealt{sieg1}, \citealt{liuj1}),
respectively. The averaged effective temperature of 
$\langle T_{\rm e}\rangle=6778$~K 
is significantly higher than 
$\langle T_{\rm e}\rangle=6433,6400,6490$~K 
(\citealt{sieg1}, \citealt{liuj1},
\citealt{barc1}). The minimal, averaged and maximal angular radii are
$(1.519,1.669,1.801)\times 10^{-10}$~rad 
corresponding to 
$R_{\rm min}=(4.46,4.90,5.29)R_\odot$, 
respectively. The following
equilibrium luminosity and effective temperature \citep{carn1} give
the position of SU Dra in a theoretical Hertzsprung-Russell diagram (HRD): 
$L_{\rm eq}=4\pi\sigma
d^2\langle\vartheta^2(\varphi) T_{\rm e}^4(\varphi)\rangle=(45.9\pm
9.3)L_{\odot}$, 
$T_{\rm eq}=\{L_{\rm eq}/4\pi\sigma
[\langle\vartheta\rangle d]^2\}^{1/4} = \langle\vartheta^2T_{\rm
  e}^4\rangle^{1/4} \langle\vartheta\rangle^{-1/2}=(6813\pm
20)\mbox{K}$. 
The magnitude averaged absolute brightness is $\langle
M_V \rangle=+0.68\pm .23$~mag.

\section{Discussion}

From the point of view of radiative transfer, the plane-parallel
approximation of the model atmospheres \citep{kuru1} could be applied
beyond doubt, because $h_0^{-1} \leq 0.03 R$ holds for the whole
pulsation cycle of an RR star. In favour of applying QSAA,
semi-quantitative arguments were that the temperature changes are very
slow even in a high amplitude RR star such as SU Dra: 
$5\mbox{K}/2500\mbox{s} <\vert\partial T_{\rm e}/\partial t\vert <
1500\mbox{K}/5000\mbox{s}$, while the characteristic times-cale of 
the radiative processes is below $200\mbox{s}$
(\citealt{oke1}, \citealt{buon1}). The effect of the variable
effective gravity can be characterized by our Condition II, which 
provides information on whether QSAA may be assumed. The limits are 
$4.5\mbox{cms}^{-2}/2500\mbox{s} 
    < \vert \partial g_{\rm e}/\partial t \vert <
    14200 \mbox{cms}^{-2}/2500\mbox{s}$. 
The upper limit comes from the rising branch of the light curve, 
when the main shock hits the atmosphere.

In general, the photometric input of the present method is
identical with that of the BW method. In order to obtain the
fundamental parameters, radial velocity data and their problematic
conversion to pulsation velocities are not necessary. On the other hand, 
$\vartheta$ and $h_0$ 
must be differentiated numerically; differential quotients are
sensitive to the non-validity of QSAA.

Our photometric and hydrodynamic considerations revealed empirical
quantitative conditions to find the phase intervals of the pulsation
when the static model atmospheres of \citet{kuru1} are satisfactory to
derive the variable and non-variable physical parameters of the
pulsating atmosphere. Outside these intervals, dynamical model
atmospheres are necessary to refine the parameters from QSAA, 
which is beyond the scope of this paper. The fundamental
parameters 
$d$, ${\cal M}$, $[M]$, $E(B-V)$ 
were determined using photometric quantities only in phases when 
the QSAA was a good approximation (i.e.~both Conditions I and II were 
satisfied). The values obtained from averaging over the entire
pulsation cycle can be considered as a first approximation only,
because QSAA was assumed in all phases regardless of it being a
good or poor approximation.  The large error of 
$L_{\rm eq}$ and
$\langle M_V \rangle$ 
originates from 
$\Delta d/d \approx .1$ 
of our best value $[\ast]$ in Table~\ref{tab2}.

To give a good impression of the accuracy of inverting the $UBV(RI)_C$
photometry to physical parameters, the comparison star BD +67
708 was used, because its colours are similar to those of SU Dra
(table 3 in \citealt{barc0}). The results are $[M]=-0.77\pm .03$,
$E(B-V)=.000$, $\vartheta=(1.913\pm .002)\times 10^{-10}$~rad, $\log
g=3.59\pm .01$, $T_{\rm e}=7505\pm 5$~K. The errors are roughly 
in the same order of magnitude as those of SU Dra when
Conditions I and II are satisfied.

\subsection{Kinematics of the atmosphere}

Figs.~\ref{fig3}(c), (d) demonstrate that significant corrections must be
added to 
${\dot R}$, ${\ddot R}$ 
if the true pulsation velocity and acceleration are required at 
$0 \le \tau < 1$. 
Moreover, the triangles show another correction: 
$-4.6\mbox{kms}^{-1} < {\bar v}(R,\varphi)
=h_0^{-1}\partial v/\partial r < 8.3\mbox{kms}^{-1}$ 
if $0.1 < \varphi < 0.6$, 
i.e.~$v(r,\varphi)=v(R,\varphi)-{\bar v}(R,\varphi)h_0(R,\varphi)(R-r)
+ \cdots$ 
is the velocity profile. In other words, even in the shock free phases,
considerable phase-dependent velocity and acceleration gradients exist
in the layers $R-h_0^{-1} < r < R$ of the line formation. 
The problem of 
${\bar v}\not=0$ 
is present in all phases, i.e.~even in the
shock free intervals, which are used in modern BW analyses
(e.g. \citealt{liuj1}).

The non-uniform motion of the outermost layers introduces
uncertainty when pulsation velocities are determined, because
the centre-of-mass velocity $v_\gamma$ must be subtracted from the
observed radial velocities. A recent exposition of the problem for
Cepheid stars is given in \citet{nard1}. By definition
$\vartheta,{\dot\vartheta},{\ddot\vartheta}$ 
refer to 
$0 \le \tau \ll 1$, 
while the variable component of the radial velocity is an
average of velocities (e.g. (\ref{2.302})) over the layers 
$R-h_0^{-1} \la r < R$ 
i.e.~$0\le \tau \la 0.3$. 
The effect on 
$v_\gamma$ and $d$ 
has not yet been studied at all. Nevertheless, the importance is obvious,
since an error of 
$1\:\mbox{kms}^{-1}$ in $v_\gamma$
results in an error 
$\Delta d/d \approx 0.1$ \citep{gaut1}. 
There is a considerable uncertainty of $v_\gamma$ in the 
literature, suggesting an error of $d$ as large as a factor of 2. 
It suffices to mention that for SU Dra
$v_\gamma=-161,-166.9\mbox{kms}^{-1}$ 
are given by \citet{oke1} and \citet{liuj1} from high dispersion 
spectra and spectral masking method CORAVEL-technique, respectively.

A wavy fine structure in the variation of $R$ is clearly
seen in Figs.~\ref{fig1}, \ref{fig3}(b). The outward motion 
starts at 
$\varphi\approx 0.45,0.74,0.94\mbox{-}0.98$
while the outward motion is reversed 
$\varphi\approx 0.25,0.55,0.78$,
i.e. the atmospheric velocity is
$v(r,t)\approx 0$ 
at these phases. The wavy fine structure is 
without doubt real, because the beginnings of the outward
motion are connected with significant maxima in 
$T_{\rm e}(\varphi),\log g_{\rm e}(\varphi)$. 
The well-known bump and hump at 
$\varphi\approx 0.74,0.94\mbox{-}0.98$ 
\citep{smit1} in the light curve are caused by the precursor and 
main shocks, respectively. At 
$\varphi\approx 0.45$, a small change is
observable in the slope of the light curve \citep{barc3}. Here, 
the existence of a shock is a new finding. It is a pre-precursor shock; 
we propose the designation {\it jump} for it. By integrating radial
velocities, authors tend to smooth out the mentioned fine
structure of motions [e.g.~\citet{liuj1} in the case of SW
And]. This practice seems to be unjustified.

The velocity of the atmosphere is 
$v(r,t)\approx 0$ 
in the interval 
$\varphi\approx 0.82\mbox{-}0.90$, 
the atmosphere is roughly at a standstill, the brightness starts 
rising, a small depression of 
$\vartheta(\varphi),\log g_{\rm e}(\varphi)$ 
is visible at 
$\varphi\approx 0.84\mbox{-}0.86$, 
and 
$T_{\rm e}(\varphi)$
is monotonic. It is not clear whether the small undulation of
$\vartheta(\varphi)$ 
around 
$(1.571\mbox{-}1.579)\times 10^{-10}$ 
is real or not, because the maximal 
$\Delta\log g_{\rm e}=0.116$
was found just at 
$\varphi=0.88$.

\subsection{$T_{\rm e}$ scale, mass, distance, absolute brightness}

Surprisingly, in comparison with previous studies, considerable
differences were found only in $\langle T_{\rm e}\rangle$,
$T_{\rm eq}$. This is due to the fact that the interpolated
$T_{\rm e}(\varphi)$ depends on $\log g_{\rm e}(\varphi)$ 
and that the information from a five colour photometry was used
in a more complex manner: an averaged value from 30 colour index pairs
was utilized. The photometry covers the whole spectrum
between 350 and 1000~nm. The use of only one colour index,
e.g.~$V-K$ solely `because of its apparent merits' \citep{liuj1},
can result in a systematic error of 
$T_{\rm e}(\varphi)$. 
It can explain the 
$\approx 350$~K difference, because in that previous work
QSAA was assumed in all phases despite violating both
Conditions I and II. An inspection of the functions 
$\{T_{\rm e}^{(i)}(\log g_{\rm e},{\rm CI}_1,{\rm CI}_2,[M],E(B-V))\}_
{i=1,2}$ derived from the tables \citep{kuru1} shows that if 
merely one colour index is to be used, the optimal choice for 
determining 
$T_{\rm e}$ 
would be 
$R_C-I_C$. 
This is because 
$R_C-I_C$ 
is almost independent of 
$\log g_{\rm e}$ 
for the actual values of 
$[M]$ and $E(B-V)$ 
of SU Dra. However, in phases violating Condition I, the sole use of 
$R_C-I_C$ would also introduce a systematic error, similar to the use of
$V-K$.

There is a remarkable decrease of $\Delta d=200\rightarrow
67\mbox{pc}$, $\Delta {\cal M}=0.10\rightarrow 0.03{\cal M}_\odot$ if
our compressible QSAA is substituted for UAA, i.e.~more physical input
is used in the frame of a one-dimensional model in space. Since
(\ref{107a}) was derived from the dynamic equation (\ref{1.100}), 
the mass determination is more accurate from this while the
distance is more uncertain. This is reflected in the shape of the
curves in Fig.~\ref{fig3}(a). The large formal error of $d$
originates from the non-separability of errors in
$\vartheta,{\dot\vartheta},{\ddot\vartheta},h_0,
\partial h_0/\partial t,\partial^2 h_0/\partial t^2$ and
$a^{\rm (dyn)}(r,t)$.

A first attempt to derive 
$d$ and ${\cal M}$ 
of an RR star from
the over-simplified version (\ref{1.102}) of (\ref{1.100}) was
described by \citet{barc1}. The present results for the distance
and mass of SU Dra differ only slightly from the previous
$d=(647\pm 16)\mbox{pc}$, ${\cal M}=(0.66\pm .03){\cal M}_\odot$. The
most probable reason of the very good coincidence is the existence of
the phase island with the good QSAA at $\varphi\approx 0.93$, just
when the atmosphere is at a standstill; consequently, (\ref{1.102}) is a
good approximation because of $v\approx 0$.

The values 
$d=640$~pc, ${\cal M}=0.47{\cal M}_\odot$ 
given by \citet{liuj1} are very close to the present ones. However, some
caution is appropriate because of the underestimated uncertainties in
their derivation. Their 
${\cal M}$ 
originates from using (\ref{1.102}), i.e.  
$C=0$, the UAA corresponding to line 3 in Table~\ref{tab2}. Furthermore, 
because of the uncertain value of
$v_\gamma$, $\Delta d/d \la 0.6$ 
is well possible.

In the BW method, the propagation of the error 
$\Delta v_\gamma=1\:\mbox{kms}^{-1}$ 
can be estimated from (\ref{2.202})
as follows. Typical radius changes of an RR star are 
$R_{\rm max}-R_{\rm min} \approx 5\times 10^5$~km 
within 
$P/2\approx\mbox{half day}$.  
The error of the radius change is 
$\Delta R \approx \Delta v_\gamma P/2 \approx 5\times 10^4$~km, 
i.e.~$\Delta R/(R_{\rm max}-R_{\rm min}) \approx 0.1$. 
By 
$d=(R_{\rm max}-R_{\rm min} \pm \Delta R) 
/(\vartheta_{\rm max}-\vartheta_{\rm min})$, 
the final error will be 
$\Delta d/d\approx 0.1$. 
This considerably exceeds
the error originating from the projection factor
(e.g. \citealt{liuj1}) converting the observed radial velocity to
pulsation velocity. Furthermore, by (\ref{2.202}) and
(\ref{1.102}), the error $\Delta d/d\approx 0.1$ propagates to
an error $\Delta {\cal M}/{\cal M} \approx 0.2$.

The lower limit of the magnitude averaged visual absolute brightness
is $M_V > +0.48$~mag if $d < 724$~pc, 
while 
$M_V=+0.26,+0.01$~mag
belong to the improbable
values 
$d=800\:\mbox{and}\:900$~pc,
respectively.

\section{Conclusions}

Observed colours and magnitudes of a spherically pulsating star have
been compared with those of static Kurucz model atmospheres to
determine fundamental parameters of the star in the frame of the
QSAA. Photometric and hydrodynamic
conditions have been formulated for the validity QSAA 
in spherically pulsating stars.
\begin{itemize}
\item[(1)] The QSAA has been
  generalized for a non-uniform, compressible atmosphere with radial
  velocity gradient. This is a step forward because the hitherto
  available UAA described the motion of
  the atmosphere by effective gravity and differentiating the sole
  parameter radius $R$ of optical depth zero with respect to time.
\item[(2)] Our combined photometric and hydrodynamic method uses the
  variation of effective gravity, angular radius, velocity, 
  acceleration in the Euler
  equation.  The complete input of the Euler equation has been derived
  from photometry and theoretical model atmospheres. Spectroscopic and
  radial velocity observations were used neither explicitly nor
  implicitly.  This is a definite advantage in comparison with the
  BW method, because less observational efforts are
  needed to get the fundamental parameters and the problematic
  conversion of observed radial velocities to pulsation velocities is
  not necessary.
\item[(3)] Concerning the interpretation of multicolour photometry,
  the inputs are identical with those of the BW
  method. The present refinements are the quantitative conditions
  whether photometry can or cannot be interpreted by static model
  atmospheres.
\item[(4)] First, the phases have been selected by the
  photometric conditions when the QSAA
  is valid. Secondly, in a number of these
  phases, the laws of mass and momentum conservation have been
  applied in the Euler formalism of hydrodynamics to determine the mass and
  distance of the star from the motion of the atmospheric layers in
  the neighbourhood of zero optical depth. Afterwards, it was
  checked whether the hydrodynamic condition of the QSAA
  was satisfied in the used phases;
  i.e. phases that satisfied the photometric condition but
  violated the hydrodynamic condition had to be excluded. {\it
    Atmospheric dynamical mass} seems to be an appropriate term to
  indicate that the mass has been derived by a method which is
  completely different from a mass derived by pulsation or evolution
  theories \citep{smit1}.
\item[(5)] As a by-product, a variation procedure has been given
  for estimating atmospheric metallicity and interstellar reddening
  toward a star from $UBV(RI)_C$ photometry. This method was
  successfully applied for the non-variable comparison star BD +67
  708, giving $[M]=-0.77\pm .03$, $E(B-V)=.00$,
  $\vartheta=(1.913\pm .002)\times 10^{-10}$~rad, $\log g=3.59\pm
  .01$, $T_{\rm e}=7505\pm 5$~K.
\item[(6)] Using the $UBV(RI)_C$ photometry of the high amplitude RRab
  star SU Dra, the following fundamental parameters have been
  found:
  \[[M]=-1.60\pm .10,\]
  \[E(B-V)=0.015\pm .010,\]
  \[d=(663\pm 67)\mbox{pc},\]
  \[R_{\rm min}=4.46R_\odot,\:R_{\rm max}=5.29R_\odot,\]
  \[{\cal M}=(0.68\pm .03){\cal M}_{\odot},\]
  \[L_{\rm eq}=(45.9\pm 9.3)L_{\odot},\]
  \[T_{\rm eq}=(6813\pm 20)\mbox{K}.\] $L_{\rm eq}$ and $T_{\rm eq}$
  are approximate values, since they originate from averaging
  over the whole pulsation cycle containing phases in which the
  QSAA is merely a first
  approximation.
\item[(7)] The internal motions of the atmosphere with respect to zero
  optical depth have been found to be significant in comparison with
  velocities and accelerations derived from the 
  UAA. From the internal motions of the atmosphere,
  some constraints have been sketched for converting observed radial
  velocities to pulsation and centre of mass velocities. Estimations
  have been given for the error propagation in a Baade-Wesselink
  analysis.
\end{itemize}

\section*{Acknowledgements}  

The author is grateful to J. M. Benk\H o, Z. Koll\'ath, \'A. K\'osp\'al,
L. Szabados
and an anonymous referee for thoughtful reading of the paper and
useful comments leading to an improved presentation of the results.

\end{document}